\documentclass[11pt,a4paper]{article}
\usepackage[hyperref]{emnlp2020}
\usepackage{times}
\usepackage{latexsym}

\usepackage{subfig}
\usepackage{amsmath}
\usepackage{multirow}
\usepackage{graphicx}
\usepackage{markdown}
\usepackage{amssymb}
\usepackage[utf8]{inputenc}
\usepackage[english]{babel}
\usepackage{amsthm}

\newcommand{\sysname}{SetConv}

\usepackage{microtype}

\aclfinalcopy 
\def\aclpaperid{1575} 

\title{\sysname{}: A New Approach for Learning from Imbalanced Data}

\author{Yang Gao$^1$, Yi-Fan Li$^1$, Yu Lin$^1$, Charu Aggarwal$^2$, Latifur Khan$^1$ \\
  $^1$The University of Texas at Dallas, USA \\
  $^2$IBM T.J. Watson, USA \\
  \texttt{\{yxg122530,yli,yxl163430,lkhan\}@utdallas.edu,charu@us.ibm.com} \\
  } 

\begin{document}
\maketitle
\begin{abstract}
For many real-world classification problems, e.g., sentiment classification, most existing machine learning methods are biased towards the majority class when the Imbalance Ratio (IR) is high. To address this problem, we propose a set convolution (SetConv) operation and an episodic training strategy to extract a single representative for each class, so that classifiers can later be trained on a balanced class distribution. We prove that our proposed algorithm is permutation-invariant despite the order of inputs, and experiments on multiple large-scale benchmark text datasets show the superiority of our proposed framework when compared to other SOTA methods. 
\end{abstract}

\section{Introduction}
\begin{figure*}[t]
\centering
\subfloat[\label{fig:training}]{\includegraphics[width =0.9\textwidth]{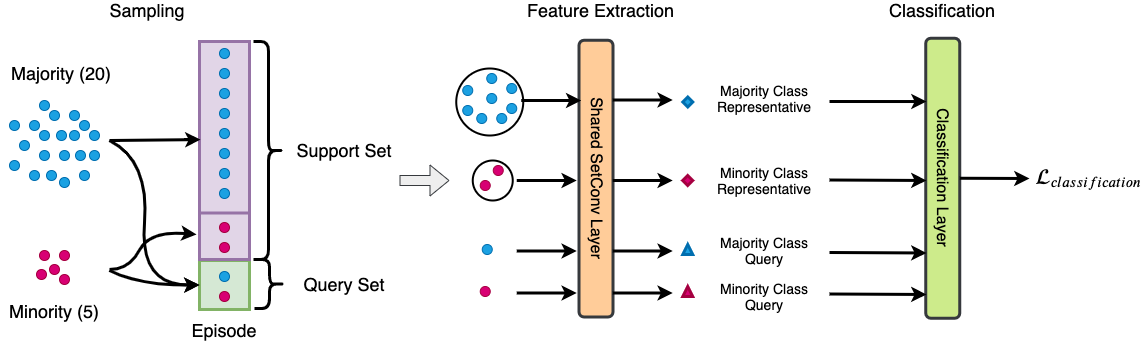}}\vfill
\subfloat[\label{fig:post_training}]{\includegraphics[width=0.9\columnwidth]{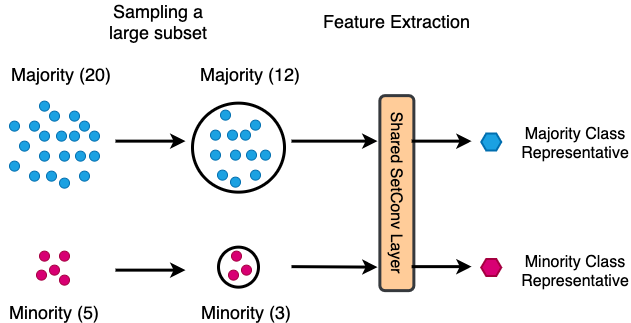}}
\subfloat[\label{fig:inference}]{\includegraphics[width = 0.9\columnwidth]{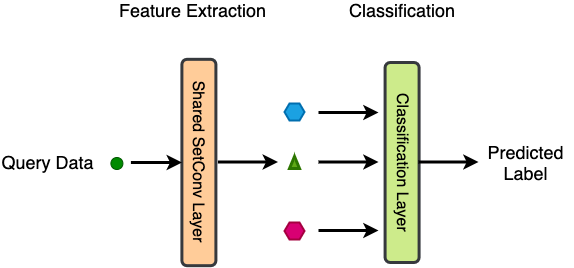}}
\caption{Overview of the proposed approach. (a) The training procedure of \sysname{}. At each iteration, \sysname{} is fed with an episode to evaluate the classification loss for model update. Each episode consists of a support set and a query set. The support set is formed by a group of samples where the imbalance ratio is preserved. The query set contains only one sample from each class. (b) The post training step of \sysname{}, which is performed only once after the main training procedure. In this step, we extract a representative for each class from the training data and will later use them for inference. Here we only perform inference using the trained model and do not update it. (c) The inference procedure of \sysname{}. Each query data is compared with every class representative to determine its label.}
\label{fig:overview}
\end{figure*}

In many real-world NLP applications, the collected data follow a skewed distribution~\cite{imagenet_cvpr09,DBLP:journals/kbs/FernandezLGJH13,DBLP:journals/jifs/YanYL17}, i.e., data from a few classes appear much more frequently than those of other classes. For example, tweets related to incidents such as shooting or fire are usually rarer compared to those about sports or entertainments. These data instances often represent objects of interest as their rareness may carry important and useful knowledge ~\cite{DBLP:journals/tkde/HeG09,DBLP:journals/pr/SunKWW07,DBLP:conf/iri/ChenS11}. However, most learning algorithms tend to inefficiently utilize them due to their disadvantage in the population~\cite{DBLP:journals/pai/Krawczyk16}. Hence, learning discriminative models with imbalanced class distribution is an important and challenging task to the machine learning community. 

Solutions proposed in previous literature can be generally divided into three categories~\cite{DBLP:journals/pai/Krawczyk16}: (1) \textit{Data-level methods} that employ under-sampling or over-sampling technique to balance the class distributions~\cite{DBLP:journals/tkde/BaruaIYM14,DBLP:journals/ml/SmithMG14,DBLP:conf/pkdd/SobhaniVM14,DBLP:journals/cai/ZhengCL15}. (2) \textit{Algorithm-level methods} that modify existing learners to alleviate their bias towards the majority classes. 
The most popular branch is the \textit{cost-sensitive} algorithms, which assign a higher cost on misclassifying the minority class instances.~\cite{DBLP:conf/ijcnn/Diaz-VicoFD18}. 
(3) \textit{Ensemble-based methods} that combine advantages of data-level and algorithm-level methods by merging data-level solutions with classifier ensembles, resulting in robust and efficient learners~\cite{DBLP:journals/tsmc/GalarFTSH12,DBLP:journals/tkde/WangMY15}. 

Despite the success of these approaches on many applications, some of their drawbacks have been observed. 
Resampling-based methods need to either remove lots of samples from the majority class or introduce a large amount of synthetic samples to the minority class, which may respectively lose important information or significantly increase the adverse correlation among samples~\cite{DBLP:conf/aaai/WuJSZY17}.
It is difficult to set the actual cost value in cost-sensitive approaches and they are often not given by expert before hand~\cite{DBLP:journals/pai/Krawczyk16}. Also, how to guarantee and utilize the diversity of classification ensembles is still an open problem in ensemble-based methods~\cite{DBLP:conf/aaai/WuJSZY17, huo2016robust}. 

In this paper, we propose a novel \textit{set convolution} (\sysname{}) operation and a new training strategy named as \textit{episodic training} to assist learning from imbalanced class distributions. 
The proposed solution naturally addresses the drawbacks of existing methods.
Specifically, \sysname{} explicitly learns the weights of convolution kernels based on the intra-class and inter-class correlations, and uses the learned kernels to  extract discriminative features from data of each class. 
It then compresses these features into a single class representative. These representatives are later applied for classification. Thus, \sysname{} helps the model to \textit{ignore sample-specific noisy information}, and \textit{focuses on the latent concept not only common to different samples of the same class but also discriminative to other classes.} On the other hand, in episodic training, we assign equal weights to different classes and do not perform resampling on data. Moreover, at each iteration during training, the model is fed with an episode formed by a set of samples where the class imbalance ratio is preserved. It encourages the model learning to \textit{extract discriminative features even when class distribution is highly unbalanced}.

Building models with \sysname{} and episodic training has several additional benefits: 

(1) \textit{Data-Sensitive Convolution.}
\ \ By utilizing \sysname{}, each input sample is associated with a set of weights that are estimated based on its relation to the minority class. This data-sensitive convolution helps the model to customize the feature extraction process for each input sample, which potentially improves the model performance.

(2) \textit{Automatic Class Balancing.}
\ \ At each iteration, no matter how many data of a class is fed into the model, \sysname{} always extracts the most discriminative information from them and compress it into a single distributed representation. Thus, the subsequent classifier, which takes these class representatives as input, always perceives a balanced class distribution.

(3) \textit{No dependence on unknown prior knowledge.}
\ \ The only prior knowledge needed in episodic training is the class imbalance ratio, which can be easily obtained from data in real-world applications.

\section{Related Work}
\subsection{Data-level Methods}
The data-level methods modifies the collection of examples by resampling techniques to balance class distributions. Existing data-level methods can be roughly classified into two categories: (1) \textit{Undersampling based methods}: this type of methods balances the distributions between the majority and minority classes by reducing majority-class samples. IHT~\cite{DBLP:journals/ml/SmithMG14} 
propose to performs undersampling based on instance hardness. On the other hand, EUS~\cite{DBLP:conf/cec/TrigueroGVCBHS15} introduce evolutionary undersampling methods to deal with large-scale classification problems. (2) \textit{Over-sampling based methods}: these methods balance the class distribution by adding samples to the minority class. SMOTE~\cite{DBLP:journals/jair/ChawlaBHK02} is the first synthetic minority oversampling technique. MWMOTE~\cite{DBLP:journals/tkde/BaruaIYM14} first identifies the hard-to-learn informative minority class samples and then uses the weighted version of these samples to generate synthetic samples. Recently, based on k-means clustering and SMOTE, KMEANS-SMOTE~\cite{DBLP:journals/corr/abs-1711-00837} is introduced to eliminate inter-class imbalance while at the same time avoiding the generation of noisy samples. 
However, for highly unbalanced data, resampling methods either discard a large amount of samples from the majority class or introduce many synthetic samples into the minority class. It leads to either the loss of important information (undersampling) or the improper increase of the adverse correlation among samples (oversampling), which will degrade the model performance~\cite{DBLP:conf/aaai/WuJSZY17}.

\subsection{Algorithm-level Methods}
Algorithm-level methods focus on modifying existing learners to alleviate their bias towards the majority classes. The most popular branch is the cost-sensitive approaches that attempt to assign a higher cost on misclassifying the minority class instances. Cost-sensitive multilayer perceptron (CS-MLP)~\cite{DBLP:journals/tnn/CastroB13} utilizes a single cost parameter to distinguish the importance of class errors. 
CLEMS~\cite{DBLP:journals/ml/HuangL17} introduces a cost-sensitive label embedding technique that takes the cost function of interest into account. 
CS-DMLP~\cite{DBLP:conf/ijcnn/Diaz-VicoFD18} is a deep multi-layer percetron model utilizing cost-sensitive learning to regularize the posterior probability distribution predicted for a given sample.
This type of methods normally requires domain knowledge to define the actual cost value, which is often hard in real-world scenarios~\cite{DBLP:journals/pai/Krawczyk16}.

\subsection{Ensemble-based Methods}
Ensemble-based methods usually combine advantages of data-level and algorithm-level methods by merging data-level solutions with classifier ensembles. A typical example is an ensemble model named as WEOB2~\cite{DBLP:journals/tkde/WangMY15} which utilizes undersampling based online bagging with adaptive weight adjustment to effectively adjust the learning bias from the majority class to the minority class. 
Unfortunately, how to guarantee and utilize the diversity of classification ensembles is still an open problem in ensemble-based methods~\cite{DBLP:conf/aaai/WuJSZY17}.

\section{Model}
\subsection{Overview}
Our goal is to develop a classification model that works well when the class distribution is highly unbalanced. For simplicity, we first consider a binary classification problem and later extend it to the multi-class scenario. As shown in Fig.~\ref{fig:training}, our model is composed of a \sysname{} layer and a classification layer. At each iteration during training, the model is fed with an \textit{episode} sampled from the training data, which is composed of a support set and a query set. The support set preserves the imbalance ratio of training data, and the query set contains only one sample from each class. Once the \sysname{} layer receives an episode, it extracts features for every sample in the episode and produces a representative for each class in the support set.
Then, each sample in the query set is compared with these class representatives in classification layer to determine its label and evaluate the classification loss for model update. We refer this training procedure as \textit{episodic training}. 

We choose episodic training due to following reasons: (1) It encourages the \sysname{} layer learning to extract discriminative features even when the class distribution of the input data is highly unbalanced. (2) Since the episodes are randomly sampled from data with significantly different configuration of support and query sets (i.e., data forming these sets vary from iteration to iteration), it requires the \sysname{} layer to capture the underlying class concepts that are common among different episodes. 

After training, a post training step is performed only once to extract a representative for each class from the training data, which will later be used for inference (Fig.~\ref{fig:post_training}). It is conducted by randomly sampling a large subset of training data (referred as $S_{post}$) and feeding them to the \sysname{} layer. \textit{Note that we only perform inference using the trained model and do not update it in this step.} We can conduct this operation because the \sysname{} layer has learned to capture the class concepts, which are insensitive to the episode configuration during training. We demonstrate it in experiments and the result is shown in Section~\ref{subsec:sensitivity}. 

The inference procedure of the proposed approach is straightforward (Fig.~\ref{fig:inference}). For each query sample, we extract its feature via the \sysname{} layer and then compare it with those class representatives obtained in post training step. The class that is most similar to the query is assigned as the predicted label.

\begin{figure}[t]
\centering
\includegraphics[width=0.9\columnwidth]{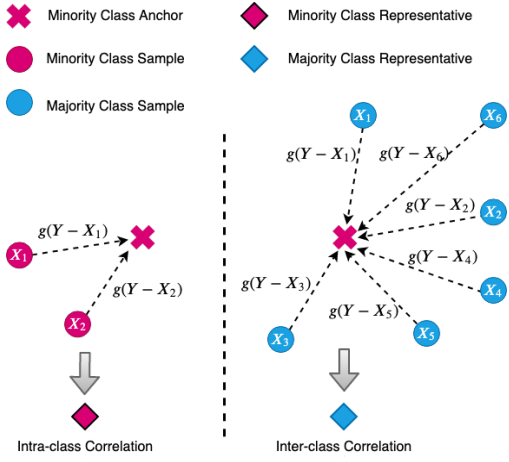}
\caption{Relations between the input samples and a pre-selected minority class anchor are used by SetConv to estimate both intra-class correlations and inter-class correlations.}
\label{fig:relative}
\end{figure}

\begin{figure*}[t]
\centering
\includegraphics[width=\textwidth]{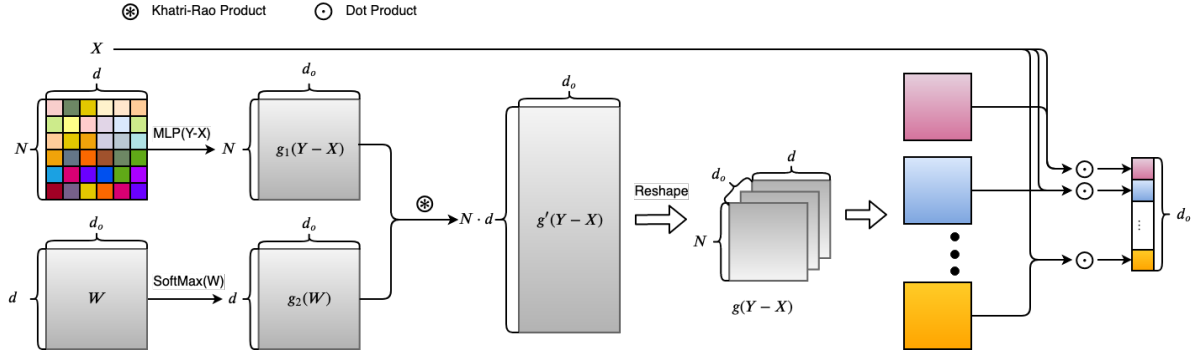}
\caption{The computation graph of the \sysname{} layer. Here $Y$ is a minority class anchor. $W \in \mathcal{R}^{d\times d_o}$ is a weight matrix to learn that records the correlation between the input and output variables. Specifically, the $i_{th}$ column of $g_2(W)$ gives the weight distribution over input features for the $i_{th}$ output feature. It is indeed a feature-level attention matrix. In addition, we estimate another data-sensitive weight matrix $g_1(Y-X)$ from the input data. The final convolution weight tensor is simply the Khatri-Rao product of $g_1(Y-X)$ and $g_2(W)$.}
\label{fig:convolution}
\end{figure*}

\subsection{\sysname{} Layer}
In many real-world applications, the minority class instances often carry important and useful knowledge that need intensive attention by the machine learning models~\cite{DBLP:journals/tkde/HeG09,DBLP:journals/pr/SunKWW07,DBLP:conf/iri/ChenS11}.

Based on this prior knowledge, we choose to design the \sysname{} layer in a way such that the \textit{feature extraction process focuses on the minority class}. We achieve it by estimating the weights of the \sysname{} layer based on the relation between the input samples and a pre-selected minority class anchor. 
This anchor can be freely determined by the user. In this paper, we adopt a simple option, i.e., \textit{average-pooling} of the minority class samples. Specifically, for each input variable, we compute its mean value across all the minority class samples in the training data. It is executable because the minority-class samples are usually limited in real-world applications\footnote{Otherwise, we may sample a subset from the minority class samples to compute the anchor.}.
As shown in Figure~\ref{fig:relative}, this weight estimation method assists the \sysname{} layer in capturing not only the intra-class correlation of the minority class, but also the inter-class correlation between the majority and minority classes.

Suppose $\mathcal{E}_t=\{\mathcal{S}_t, \mathcal{Q}_t\}$ is the episode sent to the \sysname{} layer at iteration $t$, where $\mathcal{S}_t=\big(X_{maj} \in \mathcal{R}^{N_1\times d}, X_{min} \in \mathcal{R}^{N_2\times d}\big)$ is the support set and $\mathcal{Q}_t=\big(q_{maj}\in \mathcal{R}^{1\times d}, q_{min}\in \mathcal{R}^{1\times d}\big)$ is the query set. In general, $X_{maj}$, $X_{min}$, $q_{maj}$ and $q_{min}$ can be considered as a sample set of size $N_1$, $N_2$, $1$ and $1$ respectively. For simplicity, we abstract this sample set into $X \in \mathcal{R}^{N\times d}, N\in \{N_1, N_2, 1\}$. 

Remind that the standard discrete convolution is:
\begin{equation}
    h[n]=(f\star g)[n]=\sum_{m=-M}^{m=M}f[m]g[n-m]
\end{equation}
Here, $f$ and $g$ denote the feature map and kernel weights respectively.

Similarly, in our case, we define the set convolution (\sysname{}) operation as:
\begin{equation}
\begin{split}
    h[Y] &= \frac{1}{N}\sum_{i=1}^N X_i \cdot g(Y-X_i) \\ 
    &= \frac{1}{N} \Big(X \circ g(Y-X)\Big) \\
\label{eq:set_conv}
\end{split}
\end{equation}
where $Y\in \mathcal{R}^{1\times d}$, $g(Y-X)\in \mathcal{R}^{N\times d \times d_o}$ and $h[Y]\in \mathcal{R}^{1\times d_o}$ denote the minority class anchor, kernel weights and the output embedding respectively. Here, $\circ$ is the tensor dot product operator, i.e., for every $i \in \{1, 2, \ldots, d_o\}$, we compute the dot product of $X$ and $g(Y-X)[:,:,i]$. 

Unfortunately, directly learning $g(Y-X)$ is memory intensive and computationally expensive, especially for large-scale high-dimensional data. To overcome this issue, we introduce an efficient method to approximate these kernel weights. Instead of taking $X$ as a set of $d$-dimensional samples, we stack these samples and consider it as a giant dummy sample $X'=Concat(X) \in \mathcal{R}^{1\times Nd}$. Then, Eq.~\ref{eq:set_conv} is rewritten as
\begin{equation}
    h[Y] = \frac{1}{N} \Big(X' \cdot g'(Y-X)\Big)
\end{equation}
where $g'(Y-X) \in \mathcal{R}^{Nd \times d_o}$ is the transformed kernel weights.
To efficiently compute $g'(Y-X)$, we propose to approximate it as the \textit{Khatri-Rao} product\footnote{https://en.wikipedia.org/wiki/Kronecker\_product}~\cite{DBLP:journals/corr/abs-1711-10781} of two individual components, i.e.,
\begin{equation}
\begin{split}
    g'(Y-X) &= g_1(Y-X) \circledast g_2(W)\\
            &= \text{MLP}(Y-X; \theta) \circledast \text{SoftMax}(W, 0)
\end{split}
\end{equation}
where $W\in \mathcal{R}^{d\times d_o}$ is a weight matrix that represents the correlation between input and output variables. $g_2(W)$ takes softmax over the first dimension of $W$, and is indeed a \textit{feature-level attention} matrix. The $i_{th}$ column of $g_2(W)$ provides the weight distribution over input features for the $i_{th}$ output feature. 
On the other hand, $g_1(Y-X)$ is a \textit{data-sensitive} weight matrix estimated from input data via a MLP by considering their relation to the minority class anchor.
Similar to data-level attention, $g_1(Y-X)$ helps the model customize the feature extraction process for input samples, which potentially improves the model performance. Figure~\ref{fig:convolution} shows the detailed computation graph of the \sysname{} layer.

\textbf{Discussion:}
An important property of the \sysname{} layer is \textit{\textbf{permutation-invariant}}, i.e., it is insensitive to the order of input samples. As long as the input samples are same, no matter in which order they are sent to the model, the \sysname{} layer always produces the same feature representation. Mathematically, let $\pi$ denote an arbitrary permutation matrix, we have $\sysname{}(\pi X) = \sysname{}(X)$. The detailed proof of this property is provided in the supplementary material.

\subsection{Classification}
Suppose the feature representation obtained from the \sysname{} layer for $X_{maj}$, $X_{min}$, $q_{maj}$ and $q_{min}$ in the episode are denoted by $v_{maj}^s$, $v_{min}^s$, $v_{maj}^q$ and $v_{min}^q$ respectively. The probability of predicting $v_{maj}^q$ or $v_{min}^q$ as the majority class is given by
\begin{equation}
    P(c=0|x)=\frac{\exp(x\odot v_{maj}^{s})}{\exp(x\odot v_{maj}^{s})+\exp(x\odot v_{min}^{s})}
\label{eq:probability}
\end{equation}
where $\odot$ represents the dot product operation and $x\in \{v_{maj}^q, v_{min}^q\}$.

Similarly, the probability of predicting $v_{maj}^q$ or $v_{min}^q$ as the minority class is 
\begin{equation}
    P(c=1|x)=\frac{\exp(x\odot v_{min}^{s})}{\exp(x\odot v_{maj}^{s})+\exp(x\odot v_{min}^{s})}
\end{equation}
where $x\in \{v_{maj}^q, v_{min}^q\}$.

We adopt the well-known cross-entropy loss for error estimation and use the Adam optimizer to update model.

\subsection{Extension to Multi-Class Scenario}
Extending \sysname{} for multi-class imbalance learning is straightforward. We translate the multi-class classification problem into multiple binary classification problems, i.e., we create a one-vs-all classifier for each of the $N$ classes. Specifically, for a class $c$, we treat those instances with label $y=c$ as positive and those with $y\ne c$ as negative. The anchor is hence computed based on the smaller one of the positive and negative classes. The prediction probability $P(y=c|x)$ for a given instance $x$ is computed in a similar way as Eq.~\ref{eq:probability},
\begin{equation}
    P(y=c|x)=\frac{\exp(x\odot v_{y=c}^{s})}{\exp(x\odot v_{y\ne c}^{s})+\exp(x\odot v_{y=c}^{s})}
\end{equation}
Therefore, the predicted label of the instance $x$ is $\text{argmax}_c P(y=c|x)$.

\section{Experiment}
We evaluate \sysname{} on two typical tasks, including \textit{incident detection on social media} and \textit{sentiment classification}.
\subsection{Benchmark Dataset}
\subsubsection{Incident Detection on Social Media} We conduct experiments on a real-world benchmark \textit{Incident-Related Tweet}\footnote{http://www.doc.gold.ac.uk/\%7Ecguck001/IncidentTweets/}~\cite{DBLP:journals/semweb/SchulzGJ17} (\textbf{IRT}) dataset. It contains $22,170$ tweets collected from $10$ cities, 
and allows us to evaluate our approach against geographical variations. 
The IRT dataset supports two different problem settings: binary classification and multi-class classification. In binary classification, each tweet is either ``incident-related'' or ``not incident-related''. In multi-class classification, each tweet belongs to one of the four categories including ``crash'', ``fire'', ``shooting'' and a neutral class ``not incident related''. The details of this dataset are shown in Table~\ref{tab:irt}.

\begin{table}[]
\centering
\caption{Class distribution in the IRT dataset.}
\label{tab:irt}
\resizebox{\columnwidth}{!}{%
\begin{tabular}{c|c|c|cccc}
\multirow{2}{*}{} & \multicolumn{2}{c|}{Two Classes} & \multicolumn{4}{c}{Four Classes} \\
 & Yes & No & \multicolumn{1}{c|}{Crash} & \multicolumn{1}{c|}{Fire} & \multicolumn{1}{c|}{Shooting} & No \\ \hline
Boston (USA) & 604 & 2216 & \multicolumn{1}{c|}{347} & \multicolumn{1}{c|}{188} & \multicolumn{1}{c|}{28} & 2257 \\
Sydney (AUS) & 852 & 1991 & \multicolumn{1}{c|}{587} & \multicolumn{1}{c|}{189} & \multicolumn{1}{c|}{39} & 2208 \\
Brisbane (AUS) & 689 & 1898 & \multicolumn{1}{c|}{497} & \multicolumn{1}{c|}{164} & \multicolumn{1}{c|}{12} & 1915 \\
Chicago (USA) & 214 & 1270 & \multicolumn{1}{c|}{129} & \multicolumn{1}{c|}{81} & \multicolumn{1}{c|}{4} & 1270 \\
Dublin (IRE) & 199 & 2616 & \multicolumn{1}{c|}{131} & \multicolumn{1}{c|}{33} & \multicolumn{1}{c|}{21} & 2630 \\
London (UK) & 552 & 2444 & \multicolumn{1}{c|}{283} & \multicolumn{1}{c|}{95} & \multicolumn{1}{c|}{29} & 2475 \\
Memphis (USA) & 361 & 721 & \multicolumn{1}{c|}{23} & \multicolumn{1}{c|}{30} & \multicolumn{1}{c|}{27} & 721 \\
NYC (USA) & 413 & 1446 & \multicolumn{1}{c|}{129} & \multicolumn{1}{c|}{239} & \multicolumn{1}{c|}{45} & 1446 \\
SF (USA) & 304 & 1176 & \multicolumn{1}{c|}{161} & \multicolumn{1}{c|}{82} & \multicolumn{1}{c|}{61} & 1176 \\
Seattle (USA) & 800 & 1404 & \multicolumn{1}{c|}{204} & \multicolumn{1}{c|}{153} & \multicolumn{1}{c|}{139} & 390 \\ \hline
\end{tabular}%
}
\end{table}

\begin{table}[t]
\centering
\caption{Class distribution in Amazon Review and SemiEval Datasets. }
\label{tab:sentiment_dataset}
\resizebox{0.8\columnwidth}{!}{%
\begin{tabular}{c|c|c|c}
Dataset & Negative & Positive & IR \\ \hline
Amazon-Books & 72039 & 7389 & 9.75 \\
Amazon-Electronics & 13560 & 1908 & 7.11 \\
Amazon-Movies & 12896 & 2066 & 6.24 \\
SemiEval & 39123 & 7273 & 5.38 \\ \hline
\end{tabular}%
}
\end{table}

\subsubsection{Sentiment Classification}
We conduct experiments on two large-scale benchmark datasets, including \textit{Amazon Review}\footnote{http://jmcauley.ucsd.edu/data/amazon/}~\cite{DBLP:conf/www/HeM16} and \textit{SemiEval}\footnote{http://alt.qcri.org/semeval2017/task4/index.php?id=data-and-tools}~\cite{SemEval:2017:task4}, which have been widely used for sentiment classification. Similar to MSDA~\cite{DBLP:conf/www/LiGATKT19} and SCL-MI~\cite{DBLP:conf/acl/BlitzerDP07}, we treat the amazon reviews with rating $> 3$ as positive examples, those with rating $< 3$ as negative examples, and discard the rest because their polarities are ambiguous. In addition, due to the tremendous size of Amazon Review dataset, 
we choose its $3$ largest categories, i.e., ``Books'', ``Electronics'', and ``Movies and TV'', and uniformly sample from these categories to form a subset that contains $109,858$ reviews.
This subset is sufficiently large to evaluate the effectiveness of our method. More importantly, the imbalance ratio of each category in this subset is exactly same as that in the original dataset. Details of Amazon Review and SemiEval datasets are listed in Table~\ref{tab:sentiment_dataset}.

\subsection{Baseline}
We compare our algorithm with several state-of-the-art imbalance learning methods.
\begin{itemize}
    \item \textbf{IHT}~\cite{DBLP:journals/ml/SmithMG14} (\textit{under-sampling}) is a model that performs undersampling based on instance hardness.
    \item \textbf{WEOB2}~\cite{DBLP:journals/tkde/WangMY15} (\textit{ensemble}) is an undersampling based ensemble model that effectively adjusts the learning bias from the majority class to the minority class via adaptive weight adjustment. It only supports binary classification.
    \item \textbf{KMeans-SMOTE}~\cite{DBLP:journals/corr/abs-1711-00837} (\textit{over-sampling}) is an oversampling technique that avoids the generation of noisy samples and effectively overcomes the imbalance between classes.
    \item \textbf{IML}~\cite{DBLP:conf/ijcai/0005ZJG18} (\textit{metric learning}) is a method that utilizes metric learning to explore the correlations among imbalance data and constructs an effective data space for classification.
    \item \textbf{CS-DMLP}~\cite{DBLP:conf/ijcnn/Diaz-VicoFD18} (\textit{cost-sensitive}) is a deep MLP model that utilizes cost-sensitive learning to regularize the posterior probability distribution predicted for a given sample.
\end{itemize}

\subsection{Evaluation Metric}
We use the Specificity (\textit{Spec}), Sensitivity (\textit{Sens}), $F_1$-measure ($F_1$), Geometric-Mean (\textit{G-Mean}), and the Area Under 
the receiver operating characteristic 
Curve (\textit{AUC}) to evaluate the model performance, since they are widely used in previous imbalance learning research~\cite{DBLP:conf/ijcai/0005ZJG18,DBLP:conf/ijcnn/Diaz-VicoFD18,DBLP:journals/corr/abs-1711-00837}.
The confusion matrix for multi-class classification
is shown in Table~\ref{tab:confusion_multiple}. 
In the multi-class scenario, we report the model performance for each of the minority classes because: (1) the minority classes are usually more important than the majority class in most imbalance learning problems~\cite{DBLP:journals/tkde/HeG09,DBLP:conf/iri/ChenS11}. (2) simply averaging model performance on different classes may cover model defects, especially when the class distribution is unbalanced.

\begin{table}[t]
\centering
\caption{Confusion matrix for multi-class classification problem, where $c$ denotes the class to evaluate.}
\label{tab:confusion_multiple}
\resizebox{0.9\columnwidth}{!}{%
\begin{tabular}{l|l|l}
\hline
 & Predict Label $= c$ & Predict Label $\ne c$ \\ \hline
True Label $= c$ & True Positive (TP) & False Negative (FN) \\ \hline
True Label $\ne c$ & False Positive (FP) & True Negative (TN) \\ \hline
\end{tabular}%
}
\end{table}

(1) Class-specific performance measure:
\begin{itemize}
    \item $Spec=\frac{TN}{TN+FP}$. Spec measures the model's capability to avoid false positive and finds all negative samples.
    \item $Sens=\frac{TP}{TP+FN}$. Sens measures the model's capability to avoid false negative and finds all positive samples.
\end{itemize}

(2) Overall performance measure:
\begin{itemize}
    \item $F_1=2 \cdot \frac{\textit{precision } \cdot \textit{ recall}}{\textit{precision } + \textit{ recall}}$ is the harmonic mean of $\textit{precision } = \frac{TP}{TP+FP}$ and $\textit{recall }=\frac{TP}{TP+FN}$.
    \item $\textit{G-Mean}=\sqrt{Spec \cdot Sens}$. G-Mean receives a higher value only when both Spec and Sens stay at a higher level. Thus, \textit{G-Mean} can be considered as a trade-off between \textit{Spec} and \textit{Sens}.
    \item \textit{AUC} computes the area under the ROC curve. It measures the model's capability to distinguish positive and negative classes.
\end{itemize}

In general, the model that gives higher values on these metrics is the one with better performance.

\subsection{Experiment Setup}
For all text datasets, we first pre-process each data instance via a pretrained Bert\footnote{https://github.com/huggingface/transformers}~\cite{DBLP:conf/naacl/DevlinCLT19} model to produce a 1024-dimension feature vector, which is utilized for subsequent experiments. Note that this step does not lead to any ground-truth information leakage, because Bert is trained on Wikipedia corpus in an unsupervised manner.

\begin{figure}
\centering
\includegraphics[width=\columnwidth]{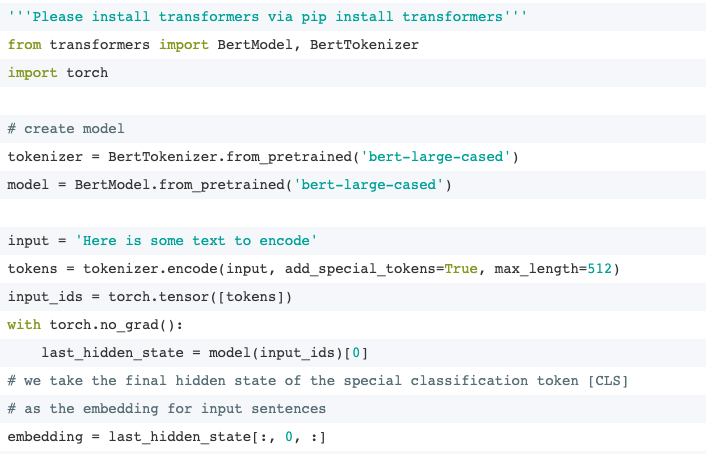}
\caption{Implementation code used to extract sentence embedding via Bert.}
\label{fig:bert}
\end{figure}

Specifically, we choose the \textit{BERT-Large, Cased (Whole Word Masking)}\footnote{https://github.com/google-research/bert} model provided by Google Research team, and take the final hidden state of the special classification token [CLS] as the embedding for any input text sequence. This process is described in Figure~\ref{fig:bert}.

After data pre-processing, we uniformly shuffle each dataset, and then divide it into development and test sets with the split ratio of 7:3. Thus, the class distribution in both development and test sets is same as that in the original dataset.
To avoid any influence of random division, we repeat the experiments $10$ times and report the average classification results.

We implement our algorithm using Python 3.7.3 and PyTorch 1.2.0 library. All baseline methods are based on code released by corresponding authors. Hyper-parameters of these baselines were set based on values reported by the authors and fine-tuned via 10-fold cross-validation on the development set. In our approach, we set the output dimension of the \sysname{} layer $d_o=128$, the size of support set $||S_{support}||=N_1+N_2=64$, the size of post-training subset $||S_{post}||=1000$, learning rate $r=0.01$, $\beta_1=0.9$ (Adam), and $\beta_2=0.999$ (Adam).  The input dimension $d$ of the \sysname{} layer is set to be the same as the dimension of input data for each dataset. The sensitivity analysis of $||S_{post}||$ is shown in Section~\ref{subsec:sensitivity}.

\begin{figure*}[t]
\centering
\includegraphics[width=\textwidth]{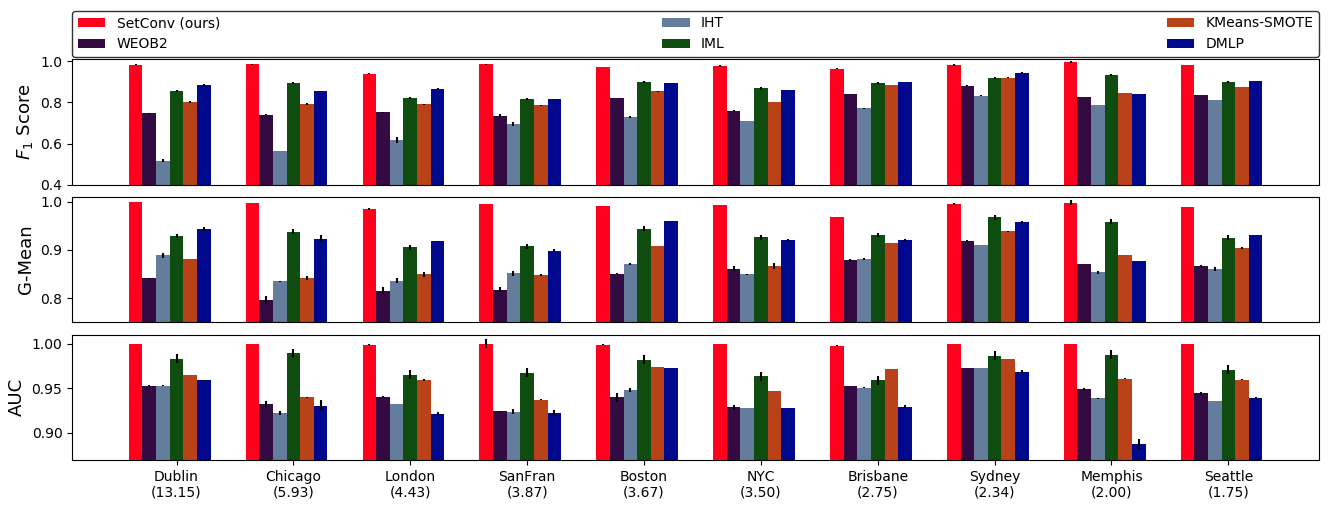}
\caption{Binary classification (incident detection) performance of competing methods on the IRT dataset. The value in the bracket indicates the imbalance ratio (IR).}
\label{fig:incident_res}
\end{figure*}

\begin{figure*}[t]
\centering
\includegraphics[width=\textwidth]{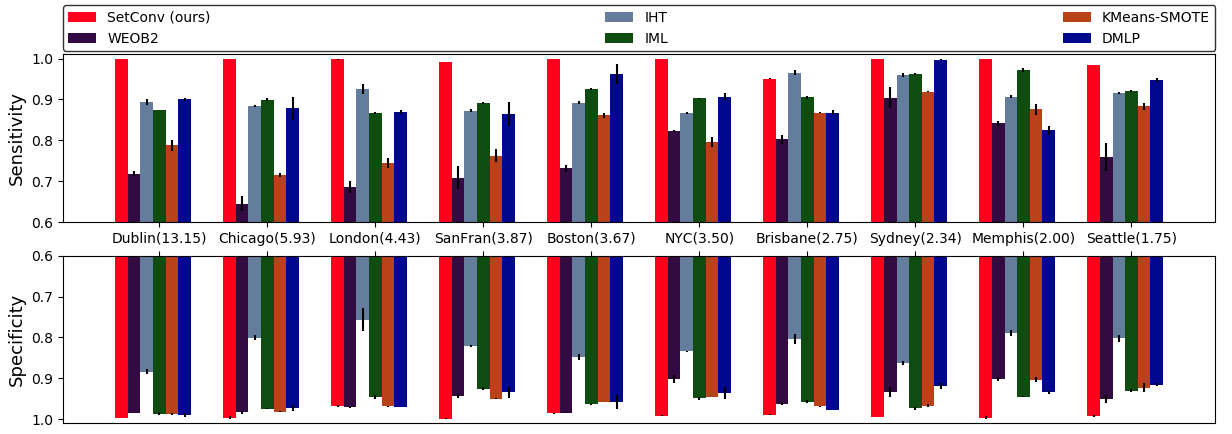}
\caption{The performance diagnosis of competing methods for binary classification. The value in the bracket indicates the imbalance ratio (IR). In contrast to baselines that are biased towards either the majority or minority class, \sysname{} performs almost equally well on both classes.}
\label{fig:incident_cls_res}
\end{figure*}

\begin{figure}[t]
\centering
\includegraphics[width=\columnwidth]{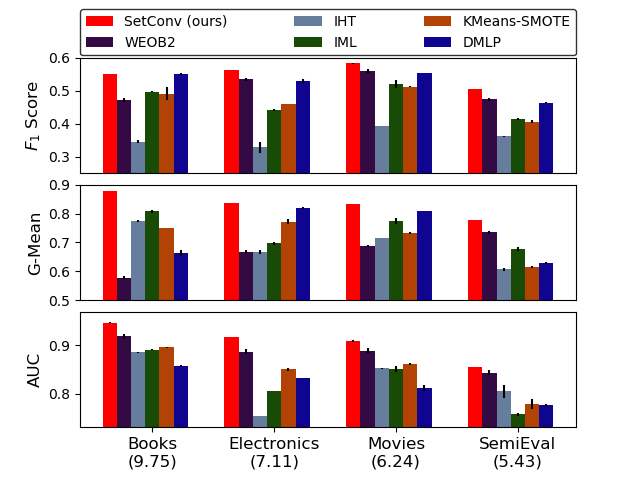}
\caption{Binary sentiment classification performance of competing methods on the Amazon Review and SemiEval datasets. The value in the bracket indicates the imbalance ratio (IR).}
\label{fig:sentiment_res}
\end{figure}

\subsection{Result}
\subsubsection{Binary Classification}
The binary classification performance of competing methods for incident detection and sentiment classification tasks are shown in Figure~\ref{fig:incident_res} and Figure~\ref{fig:sentiment_res} respectively. The results demonstrate that the proposed algorithm \textit{significantly outperforms} the competing methods in most cases and achieves the \textit{best} classification performance. Moreover, as shown in Figure~\ref{fig:incident_cls_res}, in contrast to baselines that are biased towards either the majority or minority class, the \textit{high values of specificity and sensitivity} indicate that our algorithm performs almost \textit{equally well} on both classes. That is, it not only makes few false positive predictions, but also produces few false negative predictions.
It is also observed that our method is insensitive to geographical variations. 

Our approach performs better because (1) compared to resampling based approaches, e.g., IHT and WEOB2, it makes full utilization of data via episodic training and set convolution operation, which avoids removing lot of samples from the majority class and losing important information. (2) compared to IML, \sysname{} enhances the feature extraction process by learning to extract discriminative features from a set of samples and compressing it into a single representation. It helps model to ignore sample-specific noisy information and focuses only on the latent concept common to different samples. (3) Compared to cost-sensitive approaches, e.g., CS-DMLP, episodic training assigns equal weights to both the majority and minority classes and eliminates the overhead of finding suitable cost values for different datasets. The model is forced to address class imbalance by learning to extract discriminative features during training.

\subsubsection{Multi-Class Classification}
To verify the effectiveness of the proposed algorithm in the multi-class classification scenario, we first compare it with competing methods on the IRT dataset (incident detection) and report their performance on the three minority classes, i.e., ``Fire'', ``Shooting'', and ``Crash''. Due to space limitation, we only show the results of New York City (NYC) in Table~\ref{tab:incident_detection_multiple}, although similar results have been observed for other cities. We observe that our approach significantly outperforms baseline methods by providing much higher $F_1$, G-Mean and AUC metrics. Moreover, in contrast to baseline methods, it performs almost equally well on all the three minority classes. 

In most cases, the overall classification performance of our method is superior to that of competing methods in terms of $F_1$, G-Mean and AUC metrics. Although CS-DMLP may provide better overall performance than our method in few cases, it achieves that by making many false negative predictions and missing lots of minority class samples, which is undesired in practical applications.

\begin{table}[t]
\centering
\caption{Multi-class classification performance of competing methods on the IRT-NYC dataset. $0.000$ indicates a value less than $0.0005$.
}
\label{tab:incident_detection_multiple}
\resizebox{\columnwidth}{!}{%
\begin{tabular}{|c|c|c|c|c|c|}
\hline
\multicolumn{6}{|c|}{Fire} \\ \hline
 & $F_1$ & G-Mean & AUC & Spec & Sens \\ \hline
IHT & 0.601$\pm$0.000 & 0.866$\pm$0.002 & 0.947$\pm$0.001 & 0.841$\pm$0.002 & 0.891$\pm$0.005 \\ \hline
KMeans-SMOTE & 0.831$\pm$0.005 & 0.894$\pm$0.003 & 0.967$\pm$0.001 & 0.978$\pm$0.001 & 0.818$\pm$0.005 \\ \hline
IML & 0.889$\pm$0.001 & 0.947$\pm$0.002 & 0.987$\pm$0.002 & 0.978$\pm$0.001 & 0.917$\pm$0.002 \\ \hline
CS-DMLP & 0.931$\pm$0.004 & 0.951$\pm$0.006 & 0.998$\pm$0.001 & \textbf{0.993$\pm$0.004} & 0.911$\pm$0.016 \\ \hline
SetConv (ours) & \textbf{0.972$\pm$0.002} & \textbf{0.996$\pm$0.000} & \textbf{0.999$\pm$0.000} & 0.992$\pm$0.001 & \textbf{0.999$\pm$0.001} \\ \hline
\multicolumn{6}{|c|}{Shooting} \\ \hline
 & $F_1$ & G-Mean & AUC & Spec & Sens \\ \hline
IHT & 0.333$\pm$0.001 & 0.471$\pm$0.002 & 0.984$\pm$0.001 & \textbf{0.997$\pm$0.001} & 0.222$\pm$0.002 \\ \hline
KMeans-SMOTE & 0.895$\pm$0.002 & 0.969$\pm$0.003 & 0.962$\pm$0.001 & 0.996$\pm$0.002 & 0.944$\pm$0.003 \\ \hline
IML & 0.688$\pm$0.001 & 0.780$\pm$0.002 & 0.986$\pm$0.001 & 0.996$\pm$0.001 & 0.611$\pm$0.002 \\ \hline
CS-DMLP & 0.822$\pm$0.002 & 0.910$\pm$0.029 & 0.994$\pm$0.002 & 0.995$\pm$0.002 & 0.883$\pm$0.006 \\ \hline
SetConv (ours) & \textbf{0.912$\pm$0.012} & \textbf{0.998$\pm$0.003} & \textbf{0.999$\pm$0.001} & 0.995$\pm$0.001 & \textbf{0.999$\pm$0.001} \\ \hline
\multicolumn{6}{|c|}{Crash} \\ \hline
 & $F_1$ & G-Mean & AUC & Spec & Sens \\ \hline
IHT & 0.306$\pm$0.023 & 0.762$\pm$0.019 & 0.865$\pm$0.011 & 0.755$\pm$0.020 & 0.769$\pm$0.019 \\ \hline
KMeans-SMOTE & 0.633$\pm$0.009 & 0.802$\pm$0.016 & 0.920$\pm$0.014 & 0.955$\pm$0.011 & 0.673$\pm$0.019  \\ \hline
IML & 0.662$\pm$0.002 & 0.937$\pm$0.003 & 0.959$\pm$0.001 & 0.932$\pm$0.001 & 0.942$\pm$0.003 \\ \hline
CS-DMLP & 0.702$\pm$0.054 & 0.917$\pm$0.002 & 0.969$\pm$0.013 & 0.951$\pm$0.017 & 0.885$\pm$0.019 \\ \hline
SetConv (ours) & \textbf{0.931$\pm$0.013} & \textbf{0.977$\pm$0.001} & \textbf{0.997$\pm$0.001} & \textbf{0.992$\pm$0.002} & \textbf{0.962$\pm$0.001} \\ \hline
\end{tabular}%
}
\end{table}

\subsection{Sensitivity Analysis}
\label{subsec:sensitivity}
\begin{figure}[t]
\centering
\includegraphics[width=\columnwidth]{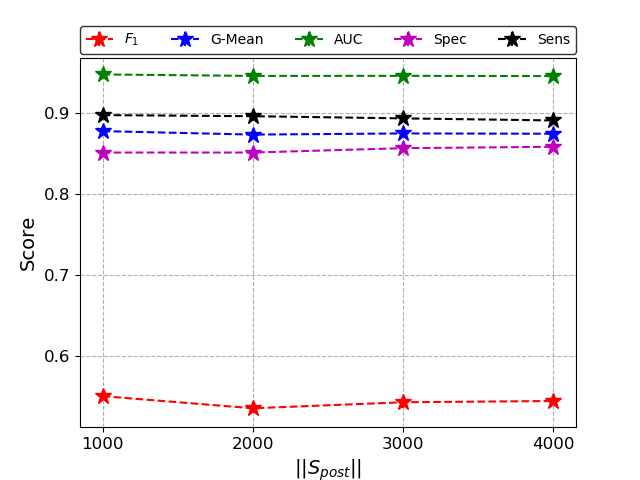}
\caption{Effect of post-training subset size ($||S_{post}||$) on classification performance.}
\label{fig:sensitivity}
\end{figure}

The main parameter in our algorithm is the size of post training subset, i.e., $||S_{post}||$. We vary $||S_{post}||$ from $1000$ to $4000$ to study its effect on the classification performance. As shown in Figure~\ref{fig:sensitivity}, our method performs stably with respect to different values of $||S_{post}||$. It demonstrates that the \sysname{} layer has learned to capture the class concepts that are common across different data samples. Thus, as long as $||S_{post}||$ is large enough, e.g., 1000, varying $||S_{post}||$ has little effect on model performance. 

\section{Conclusion}
In this paper, we propose a novel permutation-invariant \sysname{} operation and a new training strategy named as episodic training for learning from imbalanced class distributions. 
The combined utilization of them enables extracting the most discriminative features from data and automatically balancing the class distribution for the subsequent classifier. Experiment results demonstrates the superiority of our approach when compared to SOTA methods. Moreover, the proposed method can be easily migrated and applied to data of other types (e.g., images) with few modifications. 

Although the performance of \sysname{} shows its advantage
in classification, it may not be appropriate for high-dimensional sparse data. It is because the large amount of 0s in these data may lead to close-to-zero convolution kernels and limit the model's capacity for classification. Combining sparse deep learning techniques with \sysname{} is a potential solution to this issue. We leave it for future work.

\section*{Acknowledgement}
This material is based upon work supported by NSF awards DMS-1737978, DGE-2039542, OAC-1828467, OAC-1931541, DGE-1906630, and an IBM faculty award (Research).

\bibliographystyle{acl_natbib}
\bibliography{emnlp2020}

\section*{Appendix}
\subsection*{Hardware Configuration}
All experiments are performed on a server with the following hardware configuration: 
(1) 1 Intel Core i9-7920X 2.90 GHZ CPU with a total of 24 physical CPU cores
(2) 4 GeForce GTX 2080 TI GPU with 11 GB video memory
(3) 126 GB RAM. 
(4) Ubuntu 16.04 and a 4.15.0-39-generic Linux kernel.

\subsection*{Proof of Permutation Invariant Property}
An important property of the \sysname{} layer is permutation-invariant, i.e., it is insensitive to the order of input samples. As long as the input samples are same, no matter in which order they are sent to the model, the \sysname{} layer always produces the same feature representation. 

To prove it, let's consider an arbitrary permutation matrix $\pi$. Our goal is to show that $\sysname{}(\pi X) = \sysname{}(X)$. 
\begin{equation}
\begin{aligned}
    &\sysname{}(\pi X) \\
    &= \frac{1}{N} \Big(Concat(\pi X) \cdot \big[g_1(Y- \pi X) \circledast g_2(W) \big] \Big ) \\
    &= \frac{1}{N} \cdot \\ &\Big (Concat(X)E(\pi) \cdot \big[(\pi \cdot g_1(Y- X)) \circledast g_2(W) \big] \Big ) \\
    &= \frac{1}{N} \Big(X'E(\pi) \cdot E[\pi]^T \big[ g_1(Y- X) \circledast g_2(W) \big] \Big )\\
    &= \frac{1}{N} \Big( X' \cdot I \cdot \big[ g_1(Y- X) \circledast g_2(W) \big] \Big ) \\
    &= \sysname{}(X) \\
\end{aligned}
\end{equation}
Here $Concat$ is the concatenation operation which transforms a $N$-by-$d$ matrix into a $Nd$-dimensional row vector. $E(\pi)$ is the expansion operator for the permutation matrix $\pi$. For example, considering a $2$-by-$2$ permutation matrix, \[\pi = \begin{bmatrix} 0 & 1\\ 1 & 0 \end{bmatrix} \]
$E(\pi)$ is given by:
\[
E(\pi)=\begin{bmatrix} 
0 & 0 & 1 & 1\\
0 & 0 & 0 & 0\\
0 & 0 & 0 & 0\\
1 & 1 & 0 & 0
\end{bmatrix}
\]
For a toy example, $Concat(\pi X)$ is computed as below:
\[
\begin{bmatrix} 0 & 1\\ 1 & 0 \end{bmatrix} \begin{bmatrix} a & a\\ b & b \end{bmatrix} = \begin{bmatrix} b & b \\ a & a \end{bmatrix} \rightarrow \begin{bmatrix} b & b & a & a\end{bmatrix}
\]
On the other hand, $Concat(X)E(\pi)$ is given by
\[
\begin{aligned}
&Concat(X)E(\pi)\\
&=\begin{bmatrix}a & a & b & b\end{bmatrix} \begin{bmatrix} 
0 & 0 & 1 & 1\\
0 & 0 & 0 & 0\\
0 & 0 & 0 & 0\\
1 & 1 & 0 & 0
\end{bmatrix}\\
&= \begin{bmatrix} b & b & a & a \end{bmatrix}\\
&=Concat(\pi X)\\
\end{aligned}
\]
\end{document}